\documentclass[twocolumn,showpacs,prb]{revtex4}
\usepackage{graphicx}
\usepackage{dcolumn}
\usepackage{bm}

\begin{document}

\title{Band structure of new superconducting AlB$_2$-like ternary
silicides M(Al$_{0.5}$Si$_{0.5}$)$_2$ and
M(Ga$_{0.5}$Si$_{0.5}$)$_2$ (M= Ca, Sr and Ba).}
\author{I.R. Shein$^*$, N.I. Medvedeva and A.L. Ivanovskii}

\affiliation{Institute of Solid State Chemistry, Ural Branch of
the Russian Academy of Sciences 620219, Ekaterinburg, Russia}

\date{28 November 2002}

\begin{abstract}
The electronic band structures of the new superconducting (with
T$_C$ up to 7.7K) ternary silicides M(A$_{0.5}$Si$_{0.5}$)$_2$ (M=
Ca, Sr, Ba; A= Al, Ga)  in the AlB$_2$-type structure have been
investigated using the full-potential LMTO method. The
calculations showed that the trend in transition temperatures
doesn't  follow the changes in the density d-states at the Fermi
level and probably is associated with phonon-mode frequencies.\\

$^*$ E-mail: shein@ihim.uran.ru
\end{abstract}



\maketitle The discovery of superconductivity (SC) in hexagonal
AlB$_2$-like MgB$_2$ (T$_c$ $\sim$ 39K) \cite{Akimitsu} and
creation of promising materials based thereon$^{2-4}$ have
attracted a great deal of interest in related compounds
isostructural with magnesium diboride because of their potential
as a new superconductors. One of the remarkable results is the
synthesis by a floating zone and a Ar arc melting methods the
series of the new ternary layered silicides
Sr(Ga$_x$Si$_{1-x}$)$_2$ \cite{Imai1},
Ca(Al$_{0.5}$Si$_{0.5}$)$_2$ \cite{Imai2} and
(Ca,Sr,Ba)(Ga$_x$Si$_{1-x}$)$_2$ $^{7,8}$ with transition
temperatures T$_c$ ranging 3.3 - 7.7 K, except for
Ba(Al$_x$Si$_{1-x}$)$_2$. They have the AlB$_2$-type structure in
which Si and Al,Ga atoms are arranged in honeycomb sheets and
alkaline-earth metals are intercalated between them. Furthemore, a
series of the compounds (Ca,Sr,Ba)(Al$_x$Si$_{1-x}$)$_2$ were
synthesized by varying Al/Si (0.6$<$x$<$1.2) \cite{Lorenz} and the
maximum T$_c$ for these phases appears at the 1:1:1 composition.\\
Electrical resistivity and dc magnetization results$^{5-8}$
revealed that these layered silicides are a type-II
superconductors. The observed different T$_c$ of these phases
would be qualitatively attributed to the change in densities of
states at E$_F$, N(E$_F$) \cite{Imai3}. The Seebeck coefficient
measurements for (Ca,Sr,Ba)(Al$_x$Si$_{1-x}$)$_2$ \cite{Lorenz}
indicate that their carriers are predominantly electrons, in
contrast to the holes in magnesium diboride$^{1-4}$.\\
Recently, the first band structure study of the layered silicides
Sr(Ga$_x$Si$_{1-x}$)$_2$, x = 0.375, 0.5, 0.625 and
Ca(Al$_{0.5}$Si$_{0.5}$)$_2$ has been performed\cite{Shein}. It
has been shown that the SC properties may be due to the high
density of (Ca,Sr)d-states at the Fermi level. In the present work
we report the results of first-principles calculations for all
known 1:1:1 ternary compounds M(A$_{0.5}$Si$_{0.5}$)$_2$ (M=Ca,
Sr, Ba; A=Al, Ga) and analyze the band structure parameters in
association of SC properties for the isostructural and
isoelectronic compounds: Ca(Al$_{0.5}$Si$_{0.5}$)$_2$,
Sr(Al$_{0.5}$Si$_{0.5}$)$_2$, Ba(Al$_{0.5}$Si$_{0.5}$)$_2$ and
Ca(Ga$_{0.5}$Si$_{0.5}$)$_2$, Sr(Ga$_{0.5}$Si$_{0.5}$)$_2$,
Ba(Ga$_{0.5}$Si$_{0.5}$)$_2$. The band structures of the above
silicides were calculated by the scalar relativistic
full-potential LMTO method\cite{Savrasov}.
The lattice parameters used are listed in the Table 1.\\
Energy bands, total and site projected $\ell$-decomposed densities
of states (DOS, LDOS) of M(A$_{0.5}$Si$_{0.5}$)$_2$ are presented
in Figs. 1-7. Let us discuss the band structures of
M(A$_{0.5}$Si$_{0.5}$)$_2$ for example
Ca(Al$_{0.5}$Si$_{0.5}$)$_2$. The valence band (VB) for
Ca(Al$_{0.5}$Si$_{0.5}$)$_2$ includes four fully occupied bands
and has a width of about 10 eV. The quasi-core s-like band is
located in the interval from 10.0 to 7.8 eV below the Fermi level
and separated by a gap ($\sim$ 1.45 eV) from the hybrid
(Al,Si)sp-states which form four $\sigma$(2p$_{x,y}$) and two
$\pi$(p$_z$) bands, Fig. 1. The E(k) dependence for p$_{x,y}$ and
p$_z$ bands differs considerably. For 2p$_{x,y}$ like bands the
most pronounced dispersion of E(k) is observed along the direction
k$_{x,y}$ ($\Gamma$-K of the Brillouin zone (BZ)). These bands are
of the quasi two dimensional (2D) type. They form a quasi-flat
zone along k$_z$ ($\Gamma$-A). The (Al,Si)p$_{x,y}$ orbitals
participate strong covalent $\sigma$-states to form 2D honeycomb
network bonds of sp$^2$ type with the s states. The
(Al,Si)p$_z$-like bands are responsible for weaker $\pi$(p$_z$)
interactions. These 3D-type bands have the maximum dispersion in
the direction k$_z$($\Gamma$-A). The Cas,p,d-states are admixed to
p-like bands. The $\sigma$(p$_{x,y}$) and $\pi$(p$_z$) bands
intersect at the $\Gamma$ point of the BZ. It is important that
the (Al,Si)p-bands are located below E$_F$ and do not contain hole
states as well as those in non-superconducting AlB$_2$$^{2-4}$,
which is isoelectronic to the Ca(Al$_{0.5}$Si$_{0.5}$)$_2$. The
main contribution to the Ca(Al$_{0.5}$Si$_{0.5}$)$_2$ DOS in the
vicinity of the Fermi level is made by the Ca3d-states: their
constribution in N(E$_F$) is about 59 \% compared with 9 \% and
10\% for Alp- and Sip-states, respectively.\\ Let us compare the
band structures of Ca(Al$_{0.5}$Si$_{0.5}$)$_2$,
Sr(Al$_{0.5}$Si$_{0.5}$)$_2$ and Ba(Al$_{0.5}$Si$_{0.5}$)$_2$. The
most obvious consequence of the alkaline-earth metal variation (Ca
$\longrightarrow$ Sr $\longrightarrow$ Ba) is the decreasing of VB
width from 10.0 (Ca(Al$_{0.5}$Si$_{0.5}$)$_2$) to $\sim$ 9.1 eV
(Ba(Al$_{0.5}$Si$_{0.5}$)$_2$) caused by the increased cell
volume. The location and  dispersion of lowest d-bands depend from
the alkaline-earth metals. As is seen from Fig.1, (Sr,Ba)d-states
form  the nearly flat bands in the direction L-H, close to  E$_F$.
As a result for Sr(Al$_{0.5}$Si$_{0.5}$)$_2$,
Ba(Al$_{0.5}$Si$_{0.5}$)$_2$ the sharp peaks in LDOS of
(Sr,Ba)d-states hybridized with (Al,Si)ð-orbitals, appear which
are separated by a pseudogap from the bonding p-bands (Figs. 2-4).
The values of N(E$_F$) increases more than twice going from
Ca(Al$_{0.5}$Si$_{0.5}$)$_2$ to Ba(Al$_{0.5}$Si$_{0.5}$)$_2$. It
is important to note that the  increase in N(E$_F$) is due to
similtaneously growth in LDOS of valence states for all  atoms in
silicates , see Table 1.\\The band structures of
M(Al$_{0.5}$Si$_{0.5}$)$_2$ and M(Ga$_{0.5}$Si$_{0.5}$)$_2$ are
similar, Fig.1. Their differences are revealed in an increase in
the dispersion of $\sigma$-, $\pi$-bands in the A-L-H directions
an decreasein band gap (at $\sim$ 1.0-0.9 eV) between s- and
p-like bands for M(Ga$_{0.5}$Si$_{0.5}$)$_2$ compared with
M(Al$_{0.5}$Si$_{0.5}$)$_2$. The VB width of
M(Ga$_{0.5}$Si$_{0.5}$)$_{2}$ increases by $\sim$ 1.3-1.0 eV. The
change of the alkaline-earth metal in the sequence Ca
$\longrightarrow$ Sr $\longrightarrow$ Ba causes the increase of
the N(E$_F$), the M d-states make the main contribution to
near-Fermi DOS, Figs. 5-7, Table 1.\\
Thus, the band structure of the ternary AlB$_2$-like silicides is
quite different from SC MgB$_2$. As compared by magnesium
diboride, for M(A$_{0.5}$Si$_{0.5}$)$_2$ were found (i) the
filling of the bonding $p_{x,y}$-bands and the absence of $\sigma$
-holes, (ii) the increase of covalent interactions (due to
p-d-hybridization) between graphene-like (Al,Si) or (Ga,Si) sheets
and metal hexagonal layers and (iii) the principal change in the
composition of N(E$_F$), where the alkaline-earth metal
d-states make the main contributions (~ 55-60 \%, Table 1).\\
According to the experimental data$^{5-9}$: (i) in the silicides
M(Al$_{0.5}$Si$_{0.5}$)$_2$ the T$_c$ decreases monotonically when
the metal M is changed from Ca to Ba; (ii) in the silicides
M(Ga$_{0.5}$Si$_{0.5}$)$_2$ the T$_c$ changes slightly (within
range 3.9-5.1Ê) with the maximum (5.1K) for
Sr(Ga$_{0.5}$Si$_{0.5}$)$_2$, Table 1.\\
In framework of BCS theory, the T$_c$ can be estimated by the
McMillan equation T$_c$ $\approx$ $<\omega>$ exp{f($\lambda$)},
where $<\omega>$ is the averaged phonon frequency (inversely
proportional to the atomic masses), $\lambda$ is the
electron-phonon coupling constant
($\lambda$=N(E$_F$)$<I^2>$/$<M^2>$,$<I^2>$ is the electron-phonon
matrix element, the value of $<M^2>$ does not depend on mass and
is determined by force constants).\\ The values of N(E$_F$) (as
the contributions in N(E$_F$) from Md-,(Si, Al, Ga)p-states)
obtained here showed that (i) in the silicides
M(Al$_{0.5}$Si$_{0.5}$)$_2$ and M(Ga$_{0.5}$Si$_{0.5}$)$_2$
N(E$_F$) increases monotonically when the alkaline-earth metal is
changed from Ca to Ba (it is opposite to T$_c$ trend$^{5-9}$),
(ii) N(E$_F$) in the silicides M(Al$_{0.5}$Si$_{0.5}$)$_2$ are
higher than N(E$_F$) in the silicides M(Ga$_{0.5}$Si$_{0.5}$)$_2$
based on the same M, what also does not correlate with the
observed critical temperatures. Thus, the direct relationship
T$_c$ $\sim$ ~ N(E$_F$), suggested in \cite{Imai2} for the "strong
stoichiometric" compositions of silicides
M(A$_{0.5}$Si$_{0.5}$)$_2$ are impossible.\\ It may be supposed
that the main factor, determinimg the variation in  T$_c$ in a
number of isostructural and isoelectronic compounds
M(A$_{0.5}$Si$_{0.5}$)$_2$, is the change of phonon frequencies
depending from the atomic masses. Additionally,   the shape of DOS
(and the value of  N(E$_F$)) may be changed owing to disorder in
the distribution of (Al,Ge)/Si atoms in honeycomb layers. As a
result alkaline-earth metals will be in different trigonal-
prizmatic positions, that may lead to the splitting of near-Fermi
bands and to a decrease in N(E$_F$). This effect will be more
strong for Sr,Ba-containing silicides, where for  "ideal ordering"
state the N(E$_F$) is determined by narrow intensive peaks in DOS,
Figs. 3,4,6,7. The possibilty of chemical disordering and the
inhomogenity in the arc melted samples of silicides were noted by
authors of \cite{Imai2}.\\In summary, our calculations of the band
structure of all known ternary silicides with the stoichiometry
M(A$_{0.5}$Si$_{0.5}$)$_2$ showed that the Fermi level is located
in the region of a sharp DOS peak originated mainly from
alkaline-earth metal d-states with some contributions of
(Al,Ge,Si)p-orbitals. The T$_c$ behavior for the isostructural and
isoelectronic M(A$_{0.5}$Si$_{0.5}$)$_2$ cannot be identified with
the variation of N(E$_F$), and probably connected with
phonon-mode frequencies determined by atomic masses.\\
\\
Acknowledgement.\\
\\
This work was supported by the RFBR, grant 02-03-32971.


\begin{table}
\caption{. The lattice parameters (a,$\AA$, c/a [8]), total and
site-projected $\ell$-decomposed DOSs at the Fermi level
(N(E$_F$), states/eV) and transition temperatures (T$_c$, K) of
silicides M(A$_{0.5}$Si$_{0.5}$)$_2$ (M=Ca, Sr, Ba; A=Al, Ga).}
\begin{center}
\begin{tabular}{|c|c|c|c|c|c|c|}
\hline
Parameters &Ca(Al$_{0.5}$Si$_{0.5}$)$_2$&Sr(Al$_{0.5}$Si$_{0.5}$)$_2$&Ba(Al$_{0.5}$Si$_{0.5}$)$_2$&Ca(Ga$_{0.5}$Si$_{0.5}$)$_2$&Sr(Ga$_{0.5}$Si$_{0.5}$)$_2$&Ba(Ga$_{0.5}$Si$_{0.5}$)$_2$\\
\hline
a &4.1905&4.2407&4.2974&4.1201&4.1875&4.2587\\
c/a&1.0498&1.1171&1.1967&1.0777&1.1331&1.1985\\
\hline
M-s&0.028&0.061 &0.083 &0.017 & 0.034& 0.044\\
M-p&0.096 &0.138 & 0.199&0.073 &0.036 & 0.076\\
M-d& 0.663& 1.344& 1.460& 0.594&0.936 & 1.079\\
M-f&0.0 &0.0 &0.134 &0.0 & 0.0& 0.108\\
Al(Ga)-s&0.023&0.022 &0.018 &0.022 &0.017 &0.017 \\
Al(Ga)-p&0.101&0.345 &0.404 &0.104 &0.162 & 0.219\\
Al(Ga)-d&0.033&0.043 &0.038 &0.015 &0.018 &0.016 \\
Si-s&0.023&0.013 &0.009 &0.019 &0.014 & 0.013\\
Si-p&0.116&0.241 &0.237 &0.105 &0.163 & 0.164\\
Si-d&0.044&0.067 &0.066 &0.042 &0.051 & 0.051\\
\hline
Total&1.127&2.273 &2.611 &0.992 &1.431 &1.757 \\
\hline
T$_c$, K&7.7\cite{Imai1}&4.2\cite{Imai3}& $< 2^{8,9}$&4.3\cite{Imai2}&5.1$^{8,9}$& 3.9 \cite{Imai3}\\
   &7.8\cite{Lorenz}& & & & & \\
\hline
\end{tabular}
\end{center}
\end{table}

\begin{figure*}[!htb]
\vskip  0cm
\begin{tabular}{ccc}
1 & 2  & 3 \\

\includegraphics[width=5.2 cm,clip]{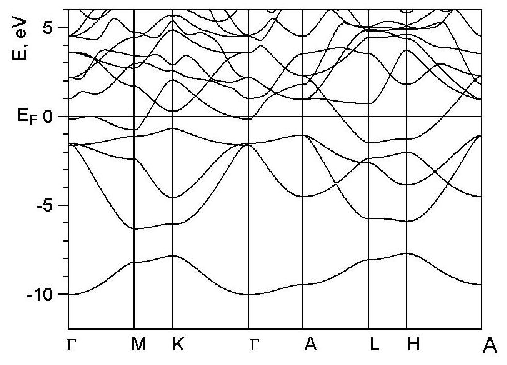}
&
\includegraphics[width=5.2 cm,clip]{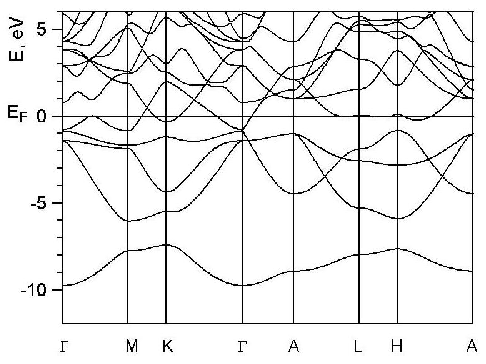}
&
\includegraphics[width=4.95 cm,clip]{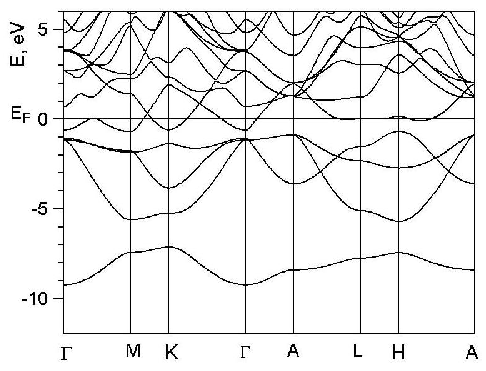} \\

4 & 5  & 6 \\

\includegraphics[width=5.2 cm,clip]{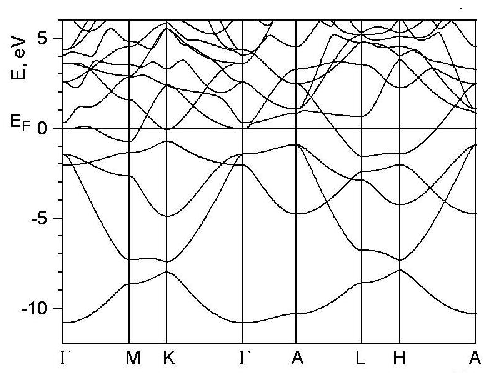}
 &
\includegraphics[width=5.2 cm,clip]{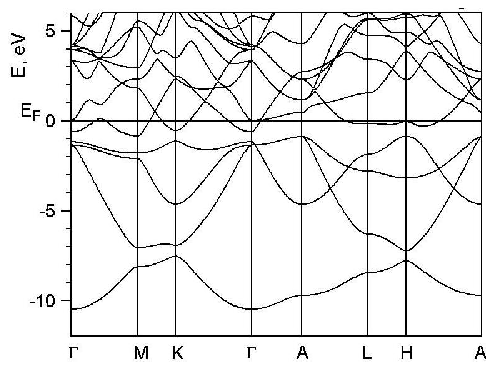}
&
\includegraphics[width=5.25 cm,clip]{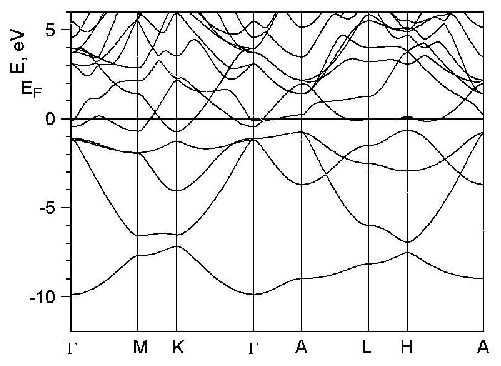} \\

\end{tabular}
\vspace{-0.02cm} \caption[a] { {\small Energy bands: 1 -
Ca(Al$_{0.5}$Si$_{0.5}$)$_2$; 2 - Sr(Al$_{0.5}$Si$_{0.5}$)$_2$; 3
- Ba(Al$_{0.5}$Si$_{0.5}$)$_2$; 4 - Ca(Ga$_{0.5}$Si$_{0.5}$)$_2$;
5 - Sr(Ga$_{0.5}$Si$_{0.5}$)$_2$; 6 -
Ba(Ga$_{0.5}$Si$_{0.5}$)$_2$. } } \label{fig:FS}
\end{figure*}

\begin{figure}[btp]
\begin{center}
\leavevmode
\includegraphics[width=2.5 in]{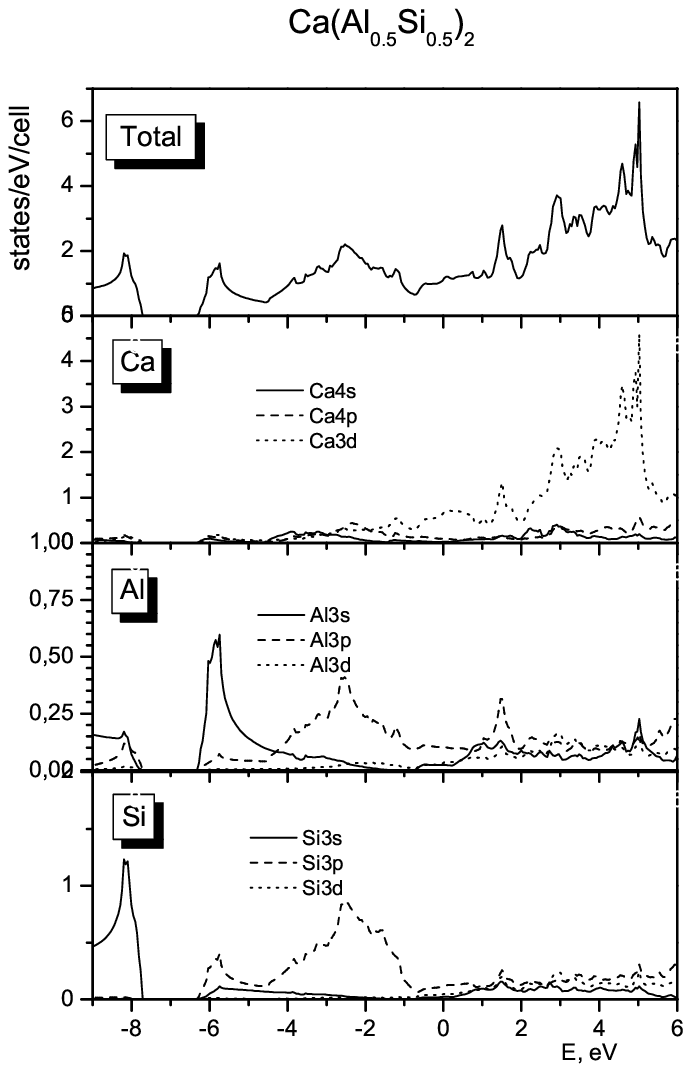}
\caption{Total and site-projected $\ell$-decomposed DOSs of
Ca(Al$_{0.5}$Si$_{0.5}$)$_2$.The energies are relative to the
Fermi level.}
\end{center}
\end{figure}

\begin{figure}[btp]
\begin{center}
\leavevmode
\includegraphics[width=2.5 in]{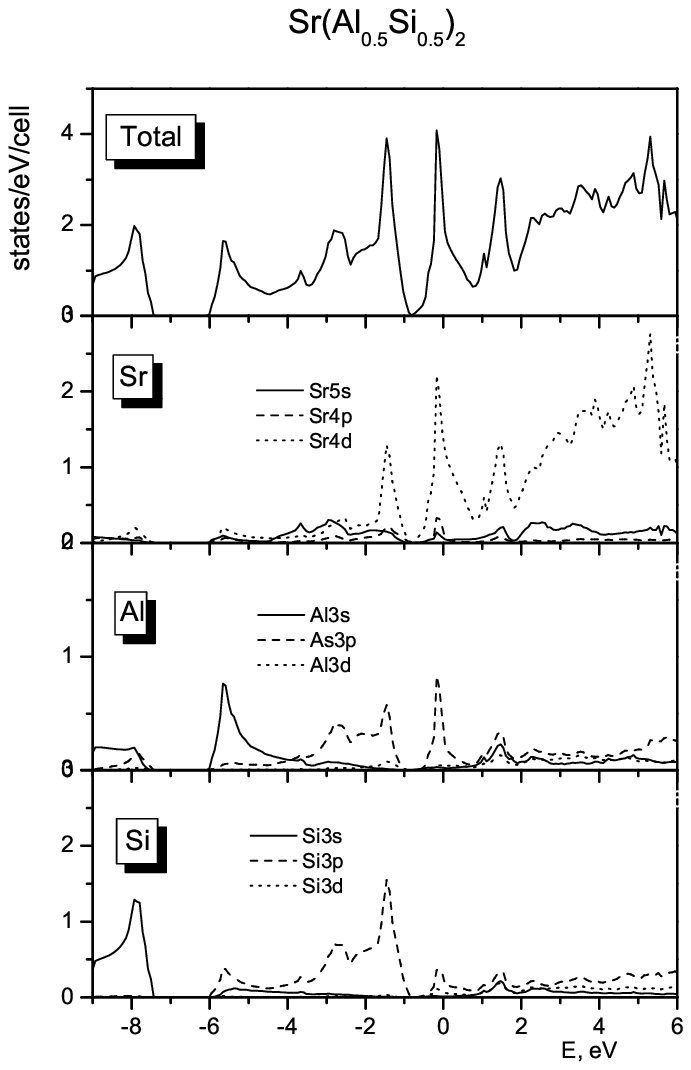}
\caption{Total and site-projected $\ell$-decomposed DOSs of
Sr(Al$_{0.5}$Si$_{0.5}$)$_2$.The energies are relative to the
Fermi level.}
\end{center}
\end{figure}

\begin{figure}[btp]
\begin{center}
\leavevmode
\includegraphics[width=2.5 in]{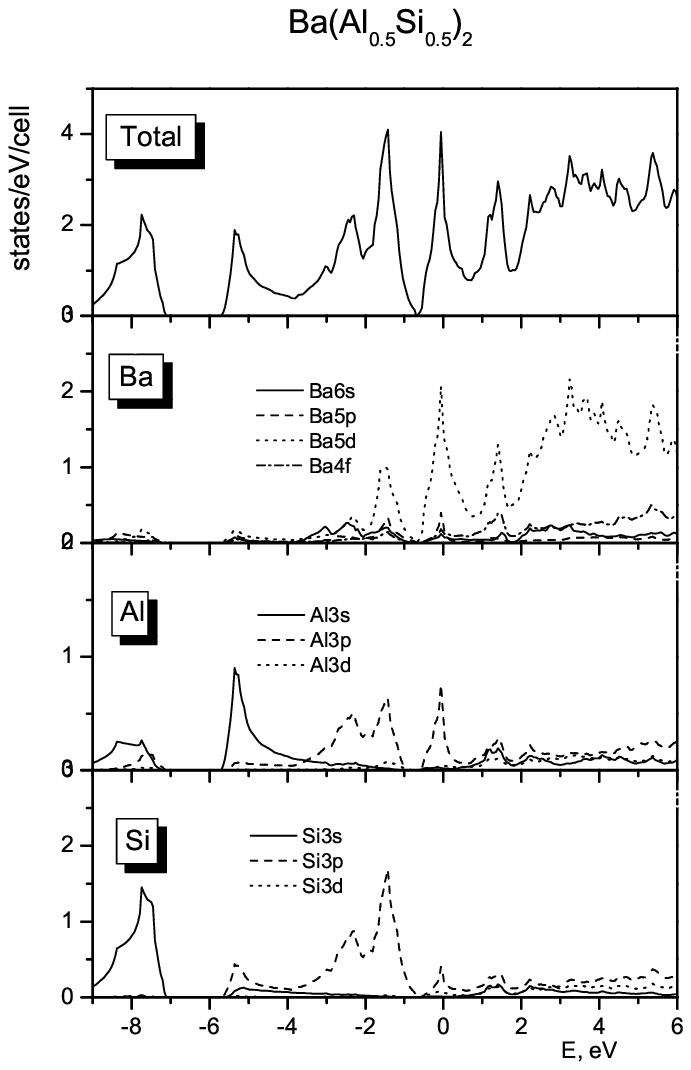}
\caption{Total and site-projected $\ell$-decomposed DOSs of
Ba(Al$_{0.5}$Si$_{0.5}$)$_2$.The energies are relative to the
Fermi level.}
\end{center}
\end{figure}

\begin{figure}[btp]
\begin{center}
\leavevmode
\includegraphics[width=2.5 in]{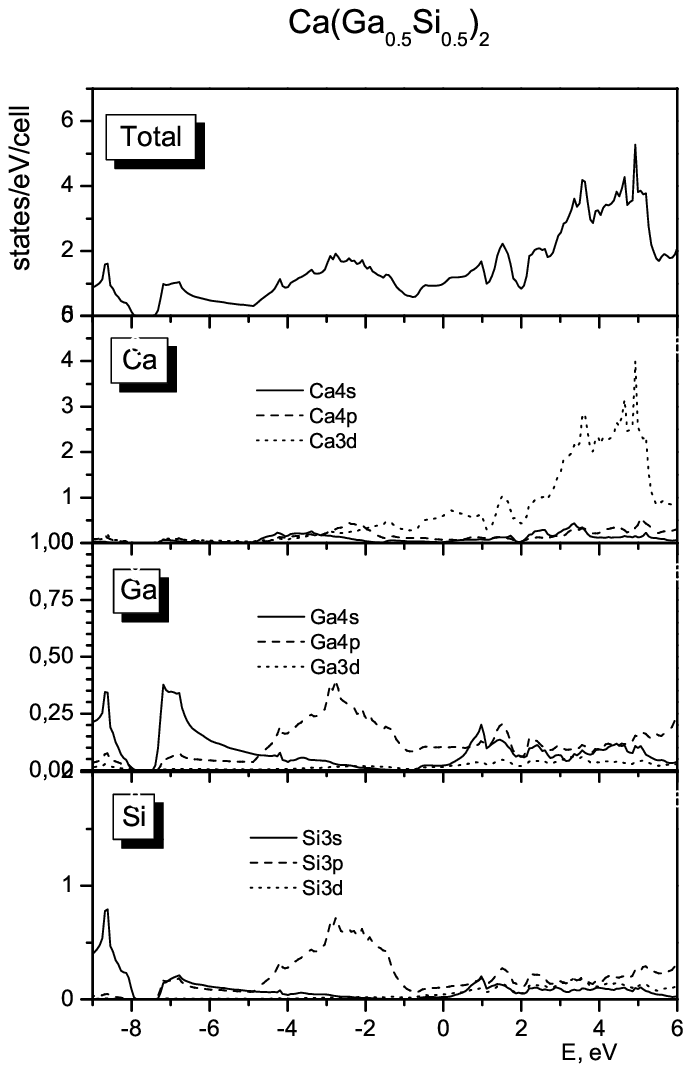}
\caption{Total and site-projected $\ell$-decomposed DOSs of
Ca(Ga$_{0.5}$Si$_{0.5}$)$_2$.The energies are relative to the
Fermi level.}
\end{center}
\end{figure}

\begin{figure}[btp]
\begin{center}
\leavevmode
\includegraphics[width=2.5 in]{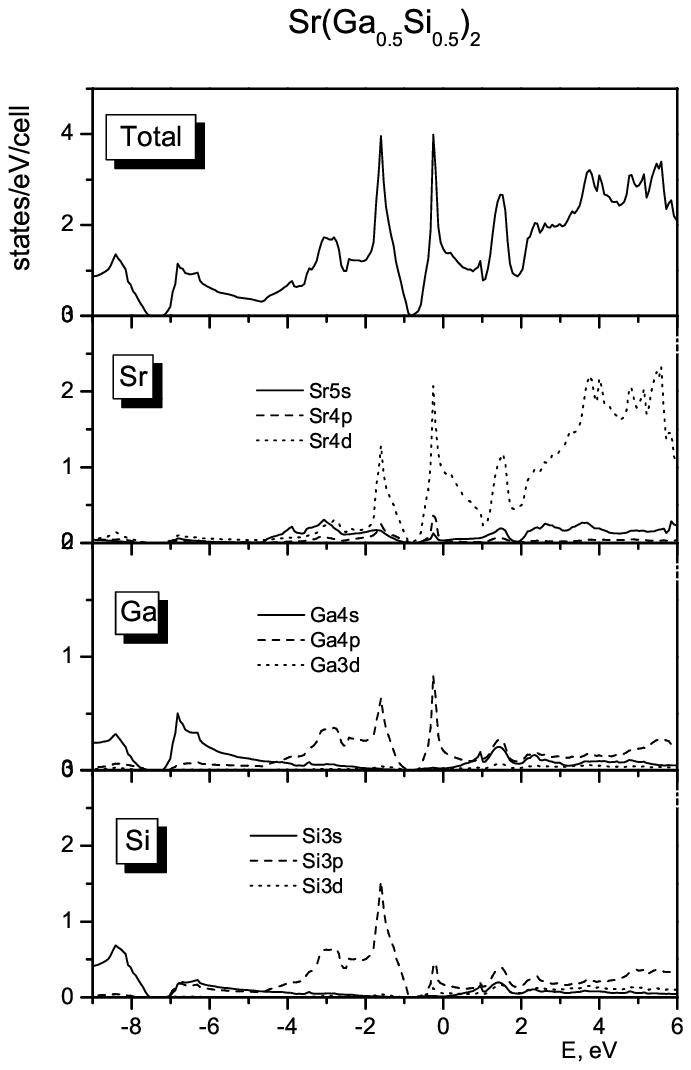}
\caption{Total and site-projected $\ell$-decomposed DOSs of
Sr(Ga$_{0.5}$Si$_{0.5}$)$_2$.The energies are relative to the
Fermi level.}
\end{center}
\end{figure}

\begin{figure}[btp]
\begin{center}
\leavevmode
\includegraphics[width=2.5 in]{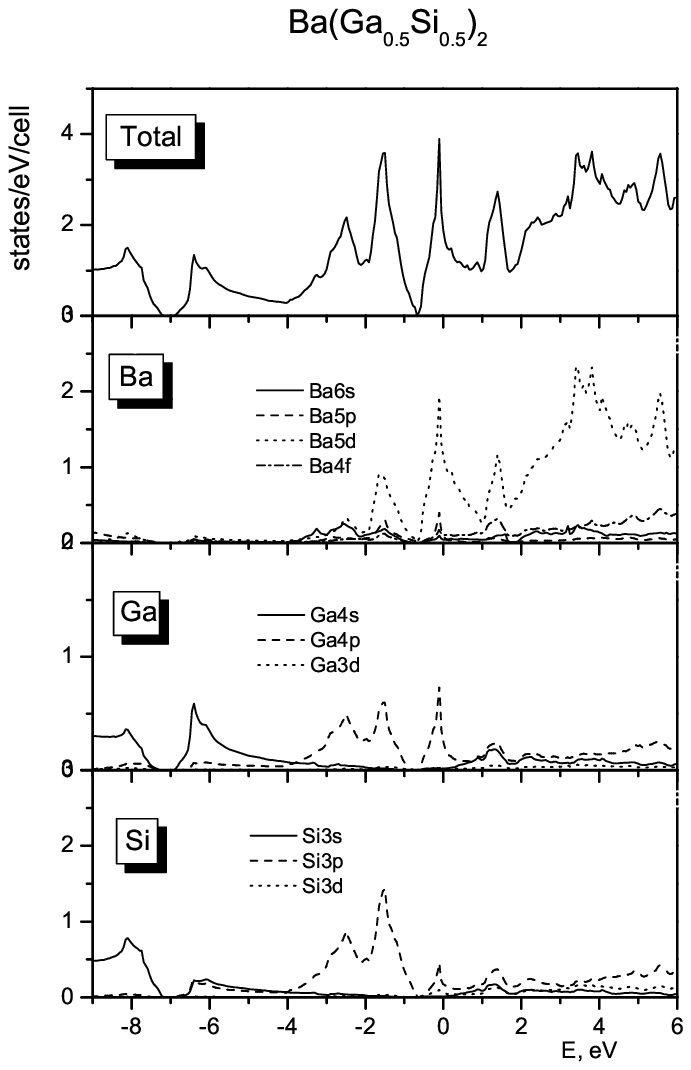}
\caption{Total and site-projected $\ell$-decomposed DOSs of
Ba(Ga$_{0.5}$Si$_{0.5}$)$_2$.The energies are relative to the
Fermi level.}
\end{center}
\end{figure}


\begin{references}

\bibitem{Akimitsu}
J. Nagamatsu, N. Nakagawa, T. Muranaka, Y. Zenitani, and J.
Akimitsu, Nature {\bf 410}, 63 (2001).

\bibitem{Ivanovskii1}
A.L. Ivanovskii, Russ. Chem. Rev., {\bf 71} 203 (2001).

\bibitem{Buzea}
C. Buzea and T. Yamashita, Supercond. Sci. Technol., {\bf 14},
R115 (2001).

\bibitem{Canfield}
P.C. Canfield and S.L. Bud'ko, Phys. World, {\bf 15} 29 (2002).

\bibitem{Imai}
M. Imai, E. Abe, J. Ye, K. Nishida, T. Kimura, K.Honma, H. Abe,
and H. Kitazawa, Phys. Rev. Letters,{\bf 87}, 077003 (2001).

\bibitem{Imai1}
M. Imai, K. Nishida, T. Kimura, and H. Abe, Appl. Phys. Letters,
{\bf 80}, 1019 (2002).


\bibitem{Imai2}
M. Imai, K. Nishida, T. Kimura, and H. Abe, Physica, {\bf C377},
96 (2002).

\bibitem{Imai3}
M. Imai, K. Nishida, T. Kimura, H. Kitazawa, H. Abe, H. Kito, and
K. Yoshii, Cond-mat/ 0210692

\bibitem{Lorenz}
B. Lorenz, J. Lenzi, J. Cmaidalka, R.L. Meng, Y.Y. Sun, Y.Y. Xue,
and C.W. Chu, Cond-mat/ 0208341

\bibitem{Shein}
I.R. Shein, V.V. Ivanovskaya, N.I. Medvedeva, and A.L. Ivanovskii,
JETP Letters, 76, 189 (2002).

\bibitem{Savrasov}
S.Y.Savrasov, Phys. Rev., {\bf B54}, 16470 (1996).

\end{references}
\end{document}